\documentclass[prlb,aps,amssym,nofootinbib,floatfix,reprint,notitlepage]{revtex4-2} 

\usepackage{xcolor}
\usepackage{amsmath,amssymb,bbold,bm}
\usepackage{graphicx}
\usepackage[normalem]{ulem}
\usepackage{cancel}
\usepackage{comment}

\usepackage[pdfpagelabels,colorlinks=true,breaklinks=true,linktocpage=true,linkcolor=red,urlcolor=blue,citecolor=blue]{hyperref}

\newcommand*{\Scale}[2][4]{\scalebox{#1}{$#2$}}
\newcommand{\sbraket}[1]{\langle #1 \rangle}
\newcommand{\be}{\begin{equation}}
\newcommand{\ben}{\begin{equation*}}
\newcommand{\ee}{\end{equation}}
\newcommand{\een}{\end{equation*}}
\newcommand{\bs}{\begin{split}}
\newcommand{\es}{\end{split}}
\newcommand{\bmx}{\begin{array}}
\newcommand{\emx}{\end{array}}

\newcommand{\bea}{\begin{eqnarray}}
\newcommand{\bean}{\begin{eqnarray*}}
\newcommand{\eea}{\end{eqnarray}}
\newcommand{\eean}{\end{eqnarray*}}
\newcommand{\dg}{^{\dagger}}
\newcommand{\dn}{^{\vphantom{\dagger}}}

\newcommand{\lr}{\leftrightarrow}

\newcommand{\bb}[1]{\mathbb{#1}}

\newcommand{\andd}{\qquad\text{and}\qquad}

\newcommand{\eps}{\epsilon}

\newcommand{\veps}{\varepsilon}

\newcommand{\sgn}[1]{{\rm sign}{#1}}
\newcommand{\pref}[1]{(\ref{#1})}

\newcommand{\intob}[1]{\int_{0}^{\beta}{#1}}

\newcommand{\im}[1]{{\rm Im}\left[ #1 \right]}
\newcommand{\tr}[1]{{\rm Tr}\left[ #1 \right]}

\newcommand{\abs}[1]{\left\vert #1 \right\vert}

\newcommand{\braket}[1]{\left\langle #1\right\rangle}

\newcommand{\mat}[1]{\left(\bmx{cc}#1\emx\right)}

\newcommand{\bw}[1]{\begin{widetext}}
\newcommand{\ew}[1]{\end{widetext}}

\newcommand{\gray}[1]{}

\newcommand{\nothing}[1]{}

\begin{document}

\title{Towards Entanglement Entropy of Random Large-N Theories}
\author{Siqi Shao}
\author{Yashar Komijani\,$^{*}$}
 \affiliation{ Department of Physics, University of Cincinnati, Cincinnati, Ohio, 45221, USA}
\date{\today}
\begin{abstract}
A large class of strongly correlated quantum systems can be described in certain large-N limits by quadratic in field actions along with self-consistency equations that determine the two-point functions. We use the replica approach and the notion of shifted Matsubara frequency to compute von Neumann and R\'enyi entanglement entropies for generic bi-partitioning of such systems. We argue that the von Neumann entropy can be computed from equilibrium spectral functions w/o partitioning, while the R\'enyi entropy requires re-calculating the spectrum in the interacting case. We demonstrate the flexibility of the method by applying it to examples of a two-site problem in presence of decoherence, and coupled Sachdev-Ye-Kitaev models.
\end{abstract}
\maketitle
\section{Introduction}
Entanglement is one of the central concepts of quantum mechanics and a notion based on which many of the modern physical phenomena are understood. The entanglement between the degrees of freedom in a region of space  A and the rest of the system $\bar{\rm A}$, is fully characterized by the so-called entanglement spectrum (ES), i.e. eigenvalues of the reduced density matrix $\rho_A={\rm Tr}_{\bar A}[\rho]$, or equivalently its various moments. Among different measures of the entanglement, R\'enyi and von Neumann entanglement entropies (EEs)
\be
\hspace{-.28cm}S^{{\rm R}_M}_A\equiv\frac{1}{1-M}\log{\rm Tr}\left[\rho_A^M\right], \quad S^{\rm vN}_A\equiv-{\rm Tr}\left[\rho_A\log\rho_A\right],\hspace{-.2cm}
\ee
are frequently used, where the latter can also be extracted from the limit $S^{\rm vN}_A=\lim_{M\to 1^+} S^{{\rm R}_M}_A$. 

It is known that the EE of typical pure states depends on the sizes of the Hilbert spaces \cite{Araki1970,Page1993}, whereas the EE of the ground state scales with the spatial extent of the regions. This is because roughly speaking, EE counts the number of entangled states; 
for gapped systems with short-range correlation an `area law' and for gapless systems with long-range correlation, a `volume law' is expected \cite{Vidal2003,Eisert2010,Fradkin2016}. 

Entanglement entropy has many important applications. For example, in 1+1 dimensional gapless systems, EE is the natural probe of the central charge of the underlying conformal field theory (CFT) \cite{Calabrese2004}. Furthermore, in 2+1 dimensional gapped systems with perimeter $L_A$, the entropy has the form $S_A^{\rm vN}=\alpha L_A-\gamma$ \cite{Zhang2012}, where $\gamma$ is a signature of topological order and can be extracted using a procedure that eliminates the extensive part  \cite{Kitaev2006,Levin2006}.

Moreover, according to eigenstate thermalization hypothesis (ETH) \cite{Deutsch1991,Srednicki1994,Dymarsky2018}, the reduced density matrix of a chaotic system in a pure state has the Boltzmann form $\rho_A\sim e^{-H_A/T_{\rm eff}}$ where $H_A$ is the Hamiltonian of \emph{detached} A part and the temperature $T_{\rm eff}$ depends on the state's energy. A somewhat unexpected example is the Laughlin state, whose ES contains the spectrum of gapless edge states that would exist if A and $\bar{\rm A}$ were physically detached \cite{Li2008}, as if due to topology and despite the gap, $\rho_A$ shares the same spectrum with $H_A$. Similar physics is present in other topological systems \cite{Yao2010,Fidkowski2010}, and is understood in terms of the relevance of the coupling between the edge states across A-$\bar{\rm A}$ border \cite{Qi2012} in the renormalization group (RG) sense. 

There are also connections to holography \cite{Ryu2006,Nishioka2009}. According to Ryu-Takayanagi conjecture, the EE of CFT$_{d+1}$ is given geometrically $S_A^{\rm vN}\propto {\cal A}_A$ by the extremal area ${\cal A}_A$ of the minimal space-like surface anchored to $A$ region and extending in the AdS$_{d+2}$ bulk. As external parameters are varied, ${\cal A}_A$ may switch from isolated surfaces to a joint surface, and this is interpreted as the formation of a wormhole. Hence, certain transitions in EE are holographically topological.

\begin{figure}[tp!]
\includegraphics[width=1\linewidth]{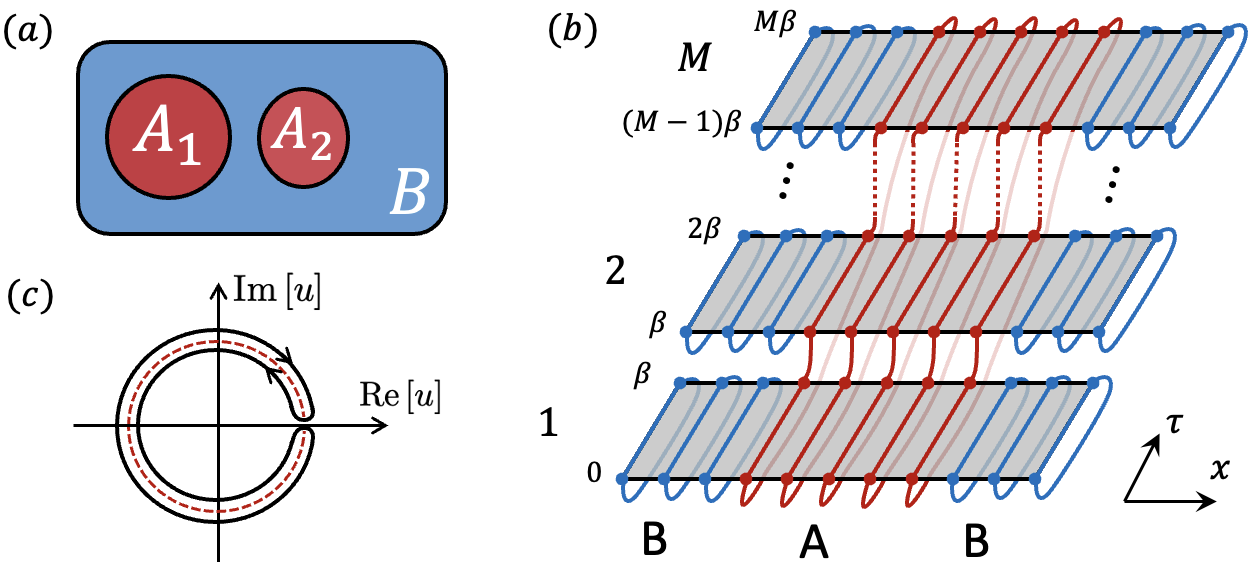}
\caption{\small (a) General bi-partite setting considered in this paper. A and B sections do not need to be simply connected. (b) The replica method for computing R\'enyi entropy. The boundary condition in the imaginary time direction for A and B sections, represented by the red/blue lines, are different. (c) The contour integral used to define von Neumann EE in the fermionic case. Bosonic case is the vertical mirror of this.}\label{fig:fig1}
\end{figure}

Despite its prevalence and important applications, the class of problems where EE can be computed are limited to non-interacting problems \cite{Bombelli1986,Casini2009,Peschel2009}, 2D CFTs \cite{Calabrese2004,Calabrese2009,Porter2016}, a number of integrable models \cite{CastroAlvaredo2008,LeBlond2019,Calabrese2020}, as well as systems amenable to quantum Monte Carlo  \cite{Grover2011,Grover2013}, exact diagonalization \cite{Regemortel2021} or density matrix renormalization. 
Here, we develop a versatile new technique that allows us to extend this list to problems which can be described by quadratic actions, e.g. models studied using static mean-field \cite{Coleman2015,Wugalter20} and dynamical large-N techniques. The latter includes random Sachdev-Ye-Kitaev (SYK) models \cite{Maldacena2016} as well as various tensor models that describe Kondo systems \cite{ofc,Komijani18,komijani2019emergent,shen2020strange,wang2020quantum,Wang2021,Drouin2021,Drouin2022,Ge2022,Wang2022,Wang2022b} and large-N theories of strange metals \cite{Esterlis2021,Guo2022}. To the best of our knowledge such a versatile technique that can be applied to all these problems was not available until now.

Previous attempts at calculating EE of these systems \cite{Gu2017,Liu2018,Haldar2020,Zhang2020,Zhang2022} have been mostly limited to 2$^{\rm nd}$ R\'enyi entropy and restricted to random SYK model, which thanks to its exact solvability and maximally chaotic behavior \cite{Maldacena2016}, have attracted considerable interest. In particular, the thermalization of the SYK and coupled-SYK models \cite{Sonner2017,Maldacena2018,Haenel2021,Kim2019,Qi2020,GarciaGarcia2021}, have been studied due to their holographic equivalence to blackholes, connected by traversable wormholes \cite{Maldacena2018,Haenel2021}. Therefore, we apply the method to study coupled-SYK models.

The rest of the paper is organized as follows. Section \ref{sec:method} is the central part of the paper where we develop our new approach to computing EE as well as comment on the role of topology. The method is then applied to various problems in section \ref{sec:examples}. We conclude in section \ref{sec:conclusion} and present some future directions. A number of appendices present supplementary information from reviewing the usual approach to non-interaction systems, the equilibrium action of the SYK model and the detailed proof of various statements made in the paper.

\section{Method}\label{sec:method}
In this section, we develop a formalism to compute EE in large-N theories which are described by quadratic action, in which the interaction is incorporated self-consistently into self-energies. We also discuss the non-interacting limit of the formalism and its connection to topology and ETH.
\subsection{Replica approach}
We consider field theories whose action ${\cal S}={\cal S}_Q+{\cal S}_C$ can be reduced into a quadratic quantum part ${\cal S}_Q[\bar\psi,\psi;\Sigma]$ in any dimension, possibly by introducing a number of dynamical constraints, and a Luttinger-Ward functional in the free energy, collected in ${\cal S}_C[G,\Sigma]$. The two parts are linked by self-consistency equations $\Sigma[G]$. We imagine dividing the system into A and ${\rm B}=\bar{\rm A}$ parts [Fig.\,\ref{fig:fig1}(a)] with (bosonic, fermionic or mixed) quantum fields $\psi_a$ and $\psi_b$, each having an arbitrary number of modes which capture the spatial extension of the region. To compute $S^{{\rm R}_M}$, we introduce $r=1\dots M$ replica of quantum fields $\psi^{(r)}(\tau)$, with imaginary-time boundary-conditions \cite{Callan1994}
\be
\psi_a^{(r)}(\beta)=\tilde\zeta^{1/M}\psi_a^{(r+1)}(0),\quad
\psi_b^{(r)}(\beta)=\tilde\zeta \psi_b^{(r)}(0),\label{eq2}
\ee
for the fields in A and B, respectively [see Fig.\,\ref{fig:fig1}b]. Here, $\tilde\zeta=\pm 1$ for bosons/fermions and we have chosen a gauge in which $\tilde\zeta$ is distributed uniformly among $\psi_a$ \cite{ftnote0}. In terms of these fields,  ${\rm Tr}[\rho_A^M]=Z_M/Z_0^{M}$ where $Z_M=e^{-N{\cal S}_C}\int{D(\bar\psi,\psi)e^{-N{\cal S}_Q}}$ has to be computed on the manifold of Fig.\,\ref{fig:fig1}(b). 
\subsection{Replica momentum}
Despite the quadratic form of the action, computing $Z_M$ is highly non-trivial due to the boundary condition \pref{eq2}. Following \cite{Casini2005} we transform both fields to the so-called \emph{replica-momentum} space,
\be
\forall p=0\dots M-1, \quad\psi^{(p)}(\tau)\equiv\frac{1}{\sqrt M}\sum_{r=1}^M\Omega ^{pr}\psi^{(r)}(\tau).
\ee
In this space the, $\psi_a$ have boundary condition $\psi^{(p)}_a(\beta)=u_p\psi_a^{(p)}(0)$ 
with $u_p\equiv\tilde\zeta^{1/M}\Omega^{-p}$ in terms $\Omega\equiv e^{2\pi i/M}$, whereas $\psi_b$ have the usual $\psi_b^{(p)}(\beta)=\tilde\zeta\psi_b^{(p)}(0)$ periodicity. For a field with periodicity $\psi(\beta)=u\psi(0)$ the Matsubara frequencies are shifted according to $u=e^{i\beta\bar\omega_n}$. Such shifted Matsubara frequency have been encountered in the perturbative calculation of 2nd R\'enyi entropy before \cite{Xu2011}, which are generalized here to arbitrary R\'enyi entropies.
The summation over shifted frequencies $\bar\omega_n$, can be done using contour integration with $n_u(z)\equiv[{ue^{\beta z}-1}]^{-1}$, and such a field has the partition sum $Z_\eps[u]\equiv[1-u^{-1}e^{\beta \eps}]^{-\zeta}$. Note that $\tilde\zeta n_{\tilde\zeta}(\omega)$ is Bose-Einstein and Fermi-Dirac distributions for $\tilde\zeta=\pm1$, respectively.

Quite generically, the quadratic action ${\cal S}_Q$ on the manifold of Fig.\,\ref{fig:fig1}(b) decouples into different $p$ sectors and using Einstein summation can be expressed as
\be
\hspace{-.3cm}{\cal S}_Q=\sum_{p=0}^{M-1}\mat{\bar \psi_{a,n} & \bar \psi_{b,m}}^{(p)}[-{\cal G}_{u_p}^{-1}]^{ab,a'b'}_{nm,n'm'}\mat{\psi_{a',n'}\\\psi_{b',m'}}^{(p)}\hspace{-.3cm}.\label{eq4}
\ee
Here $n,n'$ indices refer to shifted Matsubara frequencies $\bar\omega_{n}[u]=2\pi nT-iT\log u$, whereas $m,m'$ refer to regular bosonic/fermionic Matsubara frequencies $i\omega_m$. 
\subsection{Time-translational invariance assumption}
The R\'enyi entropy, proportional to
$\log Z_M/Z_0^M=\sum_p\log Z[u_p]/Z_0$ can be expressed as a contour integral in the complex $u$ plane [Fig.\,1(c)]
\be
\log \frac{Z_M}{Z_0^M}=\oint{\frac{du}{2\pi i}\log\Big(\frac{Z[u]}{Z_0}\Big)\partial_u\log(u^M-\tilde\zeta)}.\label{eq5b}
\ee
This enables us to extend $S_A^{{\rm R}_M}$ to non-integers values of $M$, justifying the $S^{\rm vN}_A=\lim_{\eps\to 0} S^{{\rm R}_{1+\eps}}_A$ limit. See \cite{Casini2009} for a discussion of uniqueness. Although for $u\neq \tilde\zeta$ the (imaginary) time-translational symmetry is broken \cite{Kamenev2019}, we expect it to be recovered in the $\eps\to 0$ limit and thus ${\cal G}_u(\tau_1,\tau_2)=G(\tau_1-\tau_2)+\eps\delta {\cal G}_u(\tau_1,\tau_2)$ for the Green's function. For interacting systems, this feeds into the self-energy $\Sigma=\Sigma[G]$, giving $\Sigma_u(\tau_1,\tau_2)=\Sigma(\tau_1-\tau_2)+\eps\delta\Sigma_u(\tau_1,\tau_2)$ \cite{SM}. The first observation of our paper is that since the $\eps=M-1\to 0$ limit of Eq.\,\pref{eq5b} is explicitly proportional to $\eps$, the $\eps$-correction to the self-energy is not needed to compute the von Neumann EE. Therefore, we assume that self-energy has time-translational symmetry. For non-interacting problems this is an exact statement, but for interacting large-N problems, this approximation is only valid for the von Neumann entropy. 

\subsection{Entanglement entropy formula}
Absorbing the Hamiltonian into the self-energy, the inverse Green's function in \pref{eq4} can be written as
\be
\hspace{-.34cm}\Big[{\cal G}_u^{-1}\Big]^{ab,a'b'}_{nm,n'm'}\hspace{-.5cm}=\mat{\Scale[0.8]{[i\bar\omega_n\delta^{aa'}-\Sigma^{aa'}_n]\delta_{nn'}} & \frac{1-\tilde\zeta u}{\beta}\frac{\Sigma^{ab'}_{m'}}{i\bar\omega_n-i\omega_{m'}} \\ \frac{1-\tilde\zeta u^{-1}}{\beta}\frac{\Sigma^{ba'}_m}{i\omega_m-i\bar\omega_{n'}} & \Scale[0.8]{[i\omega_m\delta^{bb'}-\Sigma^{bb'}_m]\delta_{mm'}}}\hspace{-.1cm},\hspace{-.25cm}\label{eq6b}
\ee
where we have taken advantage of time-translational symmetry of self-energies. 
The off-diagonal elements in frequency originate from the mismatch in Matsubara frequencies of $\psi_a$ and $\psi_b$ fields. However, a knowledge of equilibrium Green's function $G$ alone, is sufficient to build the ${\cal G}_u^{-1}$. See Appendix \ref{sec:action} for a derivation of Eq.\,\ref{eq6b}.

The $u$-sector partition function of action (\ref{eq4},\ref{eq6b})   is
\be
Z[u]=\det{}^{-\zeta}[(-{\cal G}_u^{-1})^{aa'}_{nn'}]\det{}^{-\zeta}[-({\cal G}_u^{BB'})_{mm'}^{-1}].\label{eq6}
\ee
	We use $\zeta=1$ for bosons, $\zeta=-1$ and $\zeta=-1/2$ for complex/real fermions and notice that $\tilde\zeta=\sgn(\zeta)$. After summation over shifted frequencies $\bar\omega_n$, and expressing the Green's function of A by its spectral representation $A^{aa'}(\omega)\equiv iG^{aa'}(z)]_{\omega-i\eta}^{\omega+i\eta}$, the ${\cal G}_u^{BB'}$ 
can be written as
\be
({\cal G}_u^{BB'})^{-1}=[G^{-1}_m]^{BB'}\delta_{mm'}-\int{\frac{dx}{2\pi}} \frac{K_u(x)\Sigma^{ba}_m A^{aa'}(x)\Sigma^{a'b'}_{m'}}{\beta(i\omega_m-x)(i\omega_{m'}-x)}\nonumber
\ee
where $K_u(\omega)\equiv(\tilde\zeta u-1){n_u(\omega)}/{n_{\tilde\zeta}(\omega)}$. Here, $G^{BB'}(z)$ with uppercase $B$ and $G^{aa'}(z)$ with lowercase $a$, are the equilibrium Green's function of the \emph{attached} ${\rm B}$ part, and \emph{detached} ${\rm A}$ part (possibly modified due to self-consistency equations), respectively. In other words, $G^{aa'}(z)$ is the inverse of the first block of ${\cal G}^{-1}_{u=\tilde\zeta}$ , but $G^{BB'}$ is the last block of the inverted matrix ${\cal G}_{u=\tilde\zeta}$. 

Using determinant shuffling technique \cite{SM} and defining $\bb 1\equiv 2\pi\delta({\omega-\omega'})\delta_{aa'}$, Eq.\,\pref{eq6} becomes
\be
Z[u]= {Z_a[u]}{Z_B}\det{}^{-\zeta}[\bb 1+\tilde\zeta K_u(\omega)\bb A_a(\omega)\bb J_A(\omega,\omega')]\label{eq7}
\ee
(see Appendix \ref{sec:eq8} for details) written in terms of 
\be
\bb J_A(\omega,\omega')\equiv\frac{1}{\beta}\sum_m\frac{\bb R_A(i\omega_m)}{(i\omega_m-\omega)(i\omega_m-\omega')}
\ee
where $R^{aa'}\equiv\Sigma^{ab} G^{BB'}\Sigma^{ba'}$.  Alternatively in terms of the attached/detached A correlators,  ${\bb R}_A=\bb G_a^{-1}\bb G_A\bb G_a^{-1}-\bb G_a^{-1}$ \cite{SM}. 
The boundary condition in imaginary-time $u$ appears in Eq.\,\pref{eq7} only via $K_u(x)$. 
We can write the determinant term as $\det{}^{-\zeta}[{n_u}/{n_{\tilde\zeta}}(\bb C-u\bb D)]$, where \cite{ftnote1}
\bea
\hspace{-.25cm}\bb C(\omega,\omega')=\tilde\zeta n_{\tilde\zeta}(\omega)\bb 1+\sqrt{\bb A_a(\omega)}\bb J_A(\omega,\omega')\sqrt{\bb A_a(\omega')},\label{eq8}
\eea
and $\bb D=\bb 1+\tilde\zeta \bb C$. Considering that $\bb J_A\to 0$ for a reference $\bb C_0$ with detached ${\rm A}$ and ${\rm B}$ parts, the system-independent thermal pre-factor can be eliminated by taking the ratio of the two determinants. Using ${\rm Tr}[\rho^M]=Z^{-M}\prod_pZ[u_p]$ and $Z=Z_aZ_B$ we finally have
\be
\tr{\rho_A^M}=\prod_p \frac{Z_a[u_p]}{Z_a}\frac{\det{}^{-\zeta}[\bb D-u_p^{-1}\bb C]}{\det{}^{-\zeta}[\bb D_0-u_p^{-1}\bb C_0]}.\label{eq9}
\ee

Eq.\,\pref{eq9} is the central result of our paper. We have succeeded to single-out the parameter $u$, characterizing the boundary condition in each sector, and express the rest in terms of equilibrium Green's functions of region A.
This enables us to evaluate the $p$-product using the identity $\prod_p\det[\bb D-u_p^{-1}\bb C]=\det[\bb D^M-\tilde\zeta\bb C^M]$. 
\subsection{Thermal part of EE}
R\'enyi entropies can be written as a sum of two terms $S^{{\rm R}_M}_A=S^{{\rm R}_M}_a+\Delta S_A^{{\rm R}_M}$. The first term is the (thermal) R\'enyi entropy of the detached  A system
\bea
S_a^{{\rm R}_M}&=&\frac{1}{1-M}[\log{Z_a(M\beta)}-M\log Z_a(\beta)],\label{eq10}
\eea
where $Z_a(\beta)=e^{-\beta F_a(\beta)}$ is the partition function of the detached A system at inverse temperature $\beta$. In the $M\to 1^+$ limit, $S_a^{\rm vN}=-dF_a/dT$ becomes the thermodynamical entropy of the detached A system. Note that $S_a(T\to 0)$, vanishes for all gapped systems, as well as most gapless systems that lack a residual $T=0$ entropy. 
\subsection{Quantum corrections}
The quantum correction to EE $\Delta S_A$, requires a diagonalization of $\bb C(\omega,\omega')$ matrix. The eigenvalues of $\bb C$ are real and positive ($c\le1$ for fermions). We define the entanglement density of states (DoS) $\Delta\rho$ as the difference $\Delta\rho(c)\equiv\rho(c)-\rho_0(c)\equiv\sum_j\delta(c-c_j)-\sum_j\delta(c-c_{j0})$ in $\bb C$ and $\bb C_0$ DoSs. $\Delta\rho(c)$ vanishes for physically detatched A and B.
Defining $c^+\equiv c+i\eta$, $\Delta\rho$ can be expressed as
\be
\Delta\rho(c)=-\frac{1}{\pi}\partial_c{\rm Im}\log\{\det[(c^+\bb 1-\bb C_0)^{-1}(c^+\bb 1-\bb C)]\}.\label{eq11}
\ee
 in terms of which, $\Delta S_A^{{\rm R}_M}=\int{dc}\Delta\rho(c)g^{{\rm R}_M}(c)$, where
\be
g^{{\rm R}_M}(c)\equiv\frac{-\zeta}{1-M}\log[(1+\tilde\zeta c)^M-\tilde\zeta c^M],\label{eq12}
\ee
and $g^{\rm vN}(c)=g^{\rm {R}_{1^+}}(c)=\zeta[(1+\tilde\zeta c)\log(1+\tilde\zeta c)-\tilde\zeta c\log(c)]$. Generally $g\ge0$, and for fermions $g\le g(1/2)=\log(2)$. 
\begin{figure}[tp]
\includegraphics[width=1\linewidth]{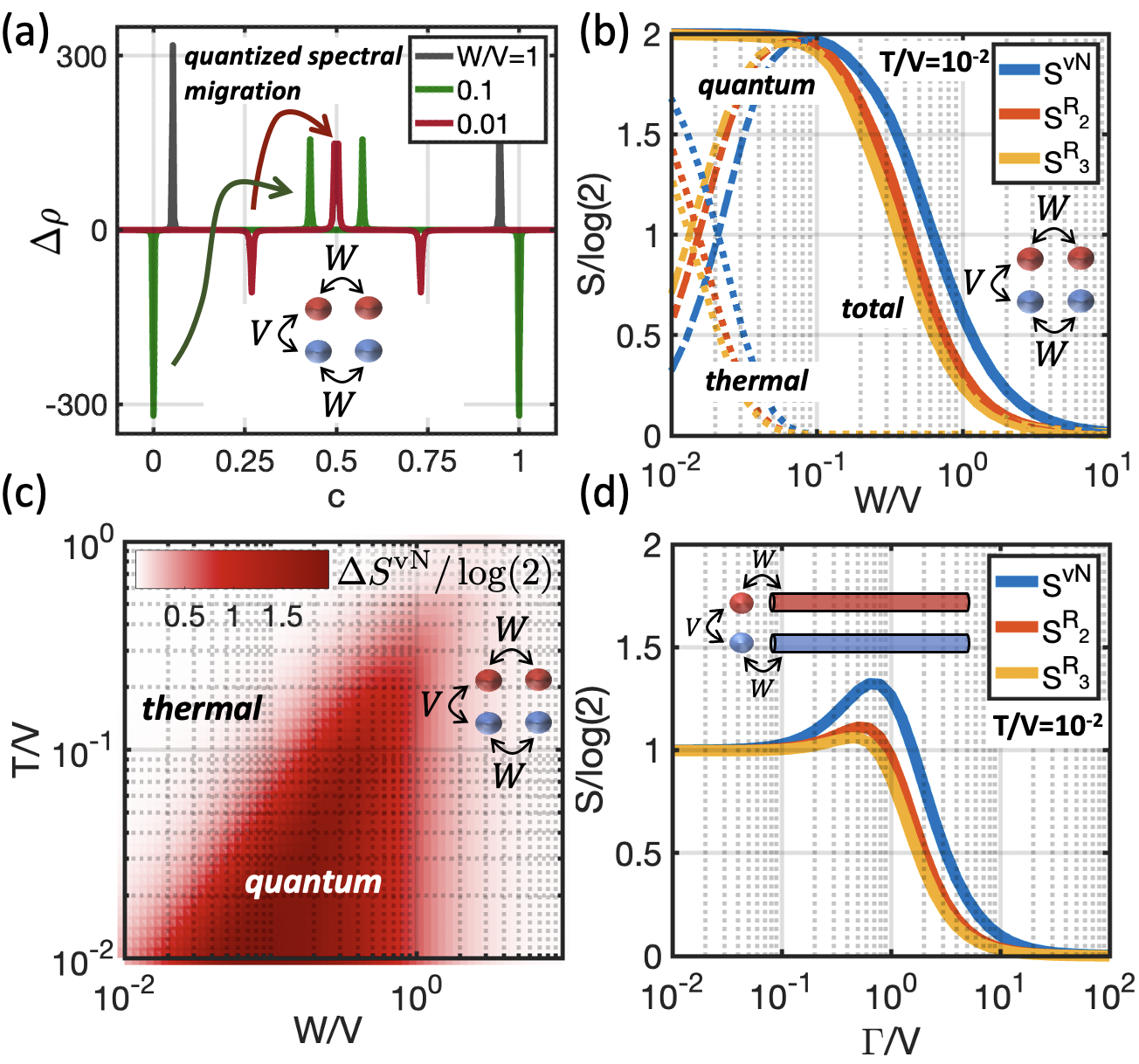}
\caption{\small Free fermions coupled to (a-c) single-site and (d) wide-band bath geometries indicated in the insets. (a) Entanglement DoS $\Delta\rho(c)$ at $T/V=10^{-2}$ for various $W/V$ ratios, show a quantized migration of positive spectrum toward $c\to 1/2$ for $T<W$, followed by a negative spectrum migration at $W<T$. Each peak has a unit area. (b) von Neumann and R\'enyi  EE as function of $W$ resolved into thermal part and quantum correction. (c) The quantum part of $S^{\rm vN}$ show its $W/T$ scaling. (d) von Neumann and R\'enyi EE as a function of $\Gamma/V$ for $\Gamma=\pi W^2/\Lambda$ in the large bandwidth $\Lambda$ limit.}\label{fig2}
\end{figure}
The matrix $\bb C$ has to be discretized and diagonalized numerically. Assuming ${\cal N}$ frequency points, $\rho$ and $\rho_0$ are each ${\cal O}({\cal N})$, but $\Delta\rho$ is an ${\cal O}(1)$ zero-mean function, independent of frequency discretization \cite{ftnote2}. The form of Eq.\,\pref{eq12} is familiar from Luttinger's theorem \cite{Seki2017}. $\Delta\rho(c)$ consists of unit-area resonances located at $c$ values where the phase of the determinant winds, corresponding to excess or deficit of an eigenvalue on top of a continuum. 

Eqs.\,(\ref{eq11}-\ref{eq12}) indicate that the $\bb C$ part of the ES can be emulated by an infinite set of auxiliary particles $\tilde\psi_{a,\omega}$ in an extra dimension \cite{Callan1994} in thermal equilibrium with occupations $\sbraket{\tilde\psi\dg_\omega\tilde\psi\dn_{\omega'}}=\bb C(\omega,\omega')$ \cite{ftnote3}. The relation $\bb C=(e^{\beta\tilde{\bb{H}}}-\tilde\zeta)^{-1}$ defines \emph{entanglement Hamiltonian} $\tilde{\bb H}$  \cite{Swingle2018}.

\subsection{Non-interacting limit and topolopgy}
In the non-interacting limit, the spectral function $\bb A^a(\omega)$ consists of a series of delta functions, which reduce the dimension of $\bb C$ to the number of modes. More importantly, $\Delta\rho<0$ contribution by $\bb C_0$ exactly cancels the thermal contribution to EE, $S_a$. In this limit $\tilde\psi_{a,\omega}\to \psi_a \delta(\omega)$, the matrix $\bb C$ represents occupation of physical particles $\psi_a$, and our formalism reduces to known results \cite{Casini2009}. See Appendix \ref{sec:non-int-limit} for details.

Generally, when A-B coupling is weaker than temperature, Eq.\,\pref{eq11} offers a perturbative expansion without the need to diagonalize $\bb C$  (See Appendix \ref{sec:perturbative}). If the A-B coupling is irrelevant in a renormalization group sense, $\Sigma^{ab}\to 0$ and $\bb J_A\to 0$ and $\Delta S$ vanishes. On the other hand, if A-B coupling is relevant, for example in presence of edge modes in the energy spectrum of detached systems, $\Sigma^{ab}\to\infty$. In this case it is justified to \emph{flatten the spectrum} \cite{Fidkowski2010} by neglecting the $k$-dependence of Green's functions involved in computing $\bb J_A(\omega,\omega')$. Writing $V^2\delta^{aa'}=\Sigma^{ab}\Sigma^{ba'}$, for each mode in A, $R^{aa'}(z)\to\delta^{aa'}V^2/(z-V^2/z)$ will have the same form as a two-site fermion problem with a coupling $V$. The latter has a zero mode in the ES and an EE of $\log(2)$. The original model has a highly degenerate zero mode, whose degeneracy is lifted by $A^a(\omega)$, resulting in a gapless mode in ES, in apparent agreement with ETH \cite{Qi2012}. 

At $T\to0$ the negative part of $\Delta\rho$ can be ignored, and  resonances can be represented by their entanglement `energies' $\veps=\log(1/c+\tilde\zeta)$. An ES gap closing and re-opening with a zero mode then indicates a topological transition in the bulk and formation of edge states. Indeed the quantum EE $\Delta S/\log(2)=\rm{nullity}(\tilde{\bb H})$ is related to the number of zero modes of $\tilde{\bb H}$, a topological invariant.

\section{Examples}\label{sec:examples}
In this section we show that the formalism developed above can be used to compute EE in large-N theories. For simplicity, we limit ourselves to two-site fermionic problems.
\subsection{Models with self-energy}
The simplest example is a system in which integrating out some internal degrees of freedom has led to a self-energy. Consider the four-site problem in a U geometry (inset of Fig.\,\ref{fig2}a) where A and B are coupled by $V$ but each are coupled by $W$ to a single-site bath, resulting in $\Sigma^{aa}(z)=\Sigma^{bb}(z)=W^2/z$. Fig.\,\ref{fig2}(b) shows EE in perfect agreement with exact diagonalization. At $W\to 0^+$, the bath sites are forced to be entangled with each other, as can be seen by a Shrieffer-Wolff produced coupling, thus $S\to 2\log(2)$.  Although the EE is constant for $W\ll V$, there is a crossover from quantum to thermal contributions as $W/T$ is varied [Fig.\,\ref{fig2}(c)]. Fig.\,\ref{fig2}(a) shows that at $W<V$ effectively two of the eigenvalues of $\bb C$ move to $c\to 1/2$, forming zero modes that increase EE to $2\log(2)$ but they are cancelled at $T<W$ by the spectral migration of $\bb C_0$ eigenvalues to zero. The EE decreases with increasing $W$, due to the entanglement monogamy. 

Our technique readily generalizes to the case where A and B are \emph{decohered} \cite{Buettiker1985} by coupling to a fermionic bath [Fig.\,\ref{fig2}(d)], an example for which many other methods fail. In the wide-band limit the self-energy can be taken to be independent of frequency, i.e. $\Sigma(\omega+i\eta)=-i\Gamma$ for both sites. The resulting EE shows no inter-bath entanglement, but an overshoot at $\Gamma\sim V$ remains.

\subsection{Models with self-consistency}
As an example of problems with self-consistency, we look at coupled-SYK models, defined as $H_0+H_{int}$ where
\be
H_0=\frac{1}{4!}\sum_{\mu=A,B}\sum^{N_\mu}_{ijkl=1}J^\mu_{ijkl}\chi^\mu_{i}\chi^\mu_{j}\chi^\mu_{k}\chi^\mu_{l}.
\ee
$H_0$ describes two copies of SYK dots. Here, $\chi^\mu_j$ are Majorana fermions and $J^\mu_{ijkl}$ are random numbers taken from a zero mean gaussian distribution (ZMGD) with the variance $J^2/N$. After disorder averaging and in the large-N$_{A,B}$ limit this model reduces to a quadratic action with the two-point function that is determined self-consistently by the self-energy $\Sigma_{\mu}(\tau)
=J^2G_{\mu}^3(\tau)$ and the Dyson equation $G_\mu^{-1}(z)=z-\Sigma_\mu(z)$. Readers are referred to \cite{Maldacena2016,Chowdhury2022} for important omitted aspects as well as Appendix \ref{sec:cSYK1} for a review of the equilibrium case. Without coupling, $\Delta S_A=0$ and thus $S_A(T)=S^{\rm SYK}_{\rm th}(T)$, which at $T\to 0$ is given by the residual entropy of a single SYK.


\begin{figure}[tp!]
\includegraphics[width=1\linewidth]{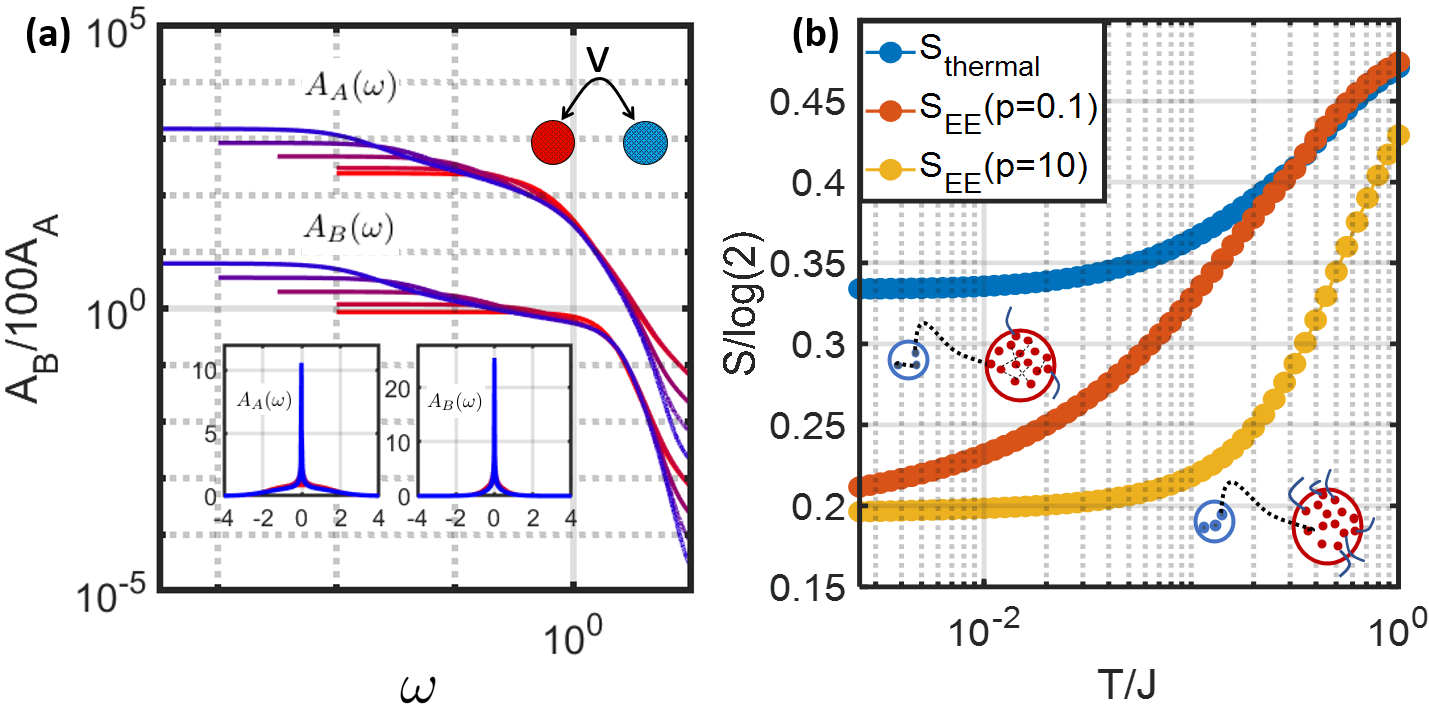}
\caption{\small Coupled-SYK model with $V/J=1$. (a) Spectral function $A_{A}$ and $A_B$ for various temperatures (T/J from $0$ to $1$), (b) von Neumann EE $S^{\rm vN}$ for different components.}\label{fig3}
\end{figure}
We now assume that the two SYK dots are connected \cite{Sohal2022} by four-fermion couplings $H=H_0+\sum_{ijkl}V_{ijkl}\chi_i^A\chi_j^A\chi_k^B\chi_l^B$, where $V_{ijkl}$ are again ZMGD with variance $V^2/N$. In this latter case, the self-consistency equations becomes $(p=N_B/N_A)$
\be
\Sigma_{aa}(\tau)
=J^2G_{AA}^3(\tau)+2V^2\sqrt{p}G^2_{BB}(\tau)G_{AA}(\tau)
\ee
and a similar equation for $\Sigma_{bb}$ with $A\lr B$ and $p\to1/p$. For $V=J$, this is a single composite SYK dot with the total number of $N=N_A+N_B$ fermions. 


A common feature of all these four-fermion coupling models is that the coupling is irrelevant.  Furthermore, $\Sigma^{ab}=\Sigma^{ba}=0$, and thus there are no quantum corrections $\Delta S_A=0$. 
{The EE is given entirely by the thermal part $S^{vN}_A=S_a$ which also includes the classical part of the action. 

Fig.\,\ref{fig3}(a) shows the spectra of A and B as well as both thermodynamical and entanglement entropy of the two systems. The same residual entropy per particle for A and B shown in Fig.\,\ref{fig3}(b)  indicates that the larger part of the coupled system is still entangled to the outside at $T=0$. In addition to that, there are some inter-subsystem entanglement as indicated by the EEs. 

\section{Conclusion}\label{sec:conclusion}

In summary, we have provided a Green's function formalism to compute ES of theories with a quadratic action through diagonalization of a single matrix built out of equilibrium functions. In this sense, our approach is different from the $\bb Z_M$ gauge theory approach taken in \cite{Iso2021,Iso2021a,Iso2021b}. This Green's function approach already simplifies the computation of entanglement entropy in specific non-interacting scenarios. 

Interactions can be treated perturbatively within this formalism. However, we argued that this formalism provides access to the von Neumann EE of large-N theories described by a Luttinger-Ward functional of two-point Green's functions. The latter includes contributions from both quantum and classical parts of the action. The focus of this work has been on the quantum part and the examples chosen are large-N models which have simple classical parts. Generalization to other examples with more complicated classical actions, e.g. \cite{Maldacena2018} is left for future. More importantly, further work is needed to investigate other systems and verify the time-translational invariant assumption that enables such an extension.

We have applied our method to a non-interacting problem with self-energy as well as the coupled SYK model. The method can be in principle applied to Kondo lattices \cite{Coleman2015,Komijani18,komijani2019emergent,Ge2022} where changes in the pattern of entanglement are shown to be playing major role in Kondo breakdown transition \cite{Wagner2018,Toldin2019,shen2020strange}. Extension to non-equllibirium steady-state as well as quench dynamics is an interesting future direction. 

Fruitful discussions with Y. Ge and R. Wijewardhana are appreciated.

\bibliography{EE}

\begin{thebibliography}{75}
\bibitem [{\citenamefont {{With real fermions}}()}]{ftnote0}%
  \BibitemOpen
  \bibfield  {author} {\bibinfo {author} {\bibnamefont {{With real
  fermions}}},\ }\href@noop {} {\bibinfo  {journal} {it is convenient to
  introduce a double copy at the beginning of the calculation and work with
  complex fermions}\ }\BibitemShut {NoStop}%
\bibitem [{\citenamefont {{We have multiplied $\bb C-u\bb D$ by $\sqrt{\bb A}$
  and its inverse from right and left}}()}]{ftnote1}%
  \BibitemOpen
\bibfield  {journal} {  }\bibfield  {author} {\bibinfo {author} {\bibnamefont
  {{We have multiplied $\bb C-u\bb D$ by $\sqrt{\bb A}$ and its inverse from
  right and left}}},\ }\href@noop {} {\bibinfo  {journal} {considering that
  $\bb A$ is positive definite and the determinant is invariant by this
  operation, to make it manifestly hermitian}\ }\BibitemShut {NoStop}%
\bibitem [{\citenamefont {{The advantage of real (as opposed to Matsubara)
  frequency formulation,}}()}]{ftnote2}%
  \BibitemOpen
\bibfield  {journal} {  }\bibfield  {author} {\bibinfo {author} {\bibnamefont
  {{The advantage of real (as opposed to Matsubara) frequency formulation,}}},\
  }\href@noop {} {\bibinfo  {journal} {is that the spectral functions decay
  rapidly beyond the bandwidth of the system. In practice $\bb
  C(\omega,\omega')$ can be replaced by $\bb C_{n,m}\equiv
  (\Delta\eps_n\Delta\eps_m)^{-1/2}\int{\bb C(\omega,\omega')d\omega
  d\omega'/2\pi}$ over arbitrary frequency ranges $\Delta\eps_{n,m}$. The
  $\Delta\rho$ is a properly regularized function. The discretization has to be
  fine enough to capture the variations of the $\bb A$ spectral function, but
  over the frequency ranges where the $\bb A$ and $\bb J$ are smooth enough, EE
  is insensitive to the choice of frequency intervals. This enables using an
  adaptive grid that emphases on fine features of spectral functions}\
  }\BibitemShut {NoStop}%
\bibitem [{\citenamefont {{The auxiliary particles have the same algebra as
  original particles}}()}]{ftnote3}%
  \BibitemOpen
  \bibfield  {author} {\bibinfo {author} {\bibnamefont {{The auxiliary
  particles have the same algebra as original particles}}},\ }\href@noop {}
  {\bibinfo  {journal} {namely,
  $\tilde\psi\dn_{a\omega}\tilde\psi\dg_{a'\omega'}-\tilde\zeta\tilde\psi\dg_{a'\omega'}\tilde\psi\dn_{a\omega}=\delta_{aa'}\delta_{\omega\omega'}$}\
  }\BibitemShut {NoStop}%
\end{thebibliography}

\appendix
\clearpage

\section*{Appendix}
This appendix contains further details and extended proof of various statements in the paper.
\subsection{Non-interacting case}\label{sec:casini}
For the sake of completeness, we remind ourselves of the known non-interacting results \cite{Peschel2009, Casini2009}.

\emph{Bosons} - The reduced density matrix is
\be
\rho_A=\frac{e^{-H_A}}{Z_A}\nonumber
\ee 
where $H_A=\sum_{mn} \eps_{mn}a_m^{\dagger}a_n$ is the entanglement Hamiltonian and $Z_A=\prod_k ({1-e^{\eps_k}})^{-1}$. We define the correlators
\be
C_{mn}\equiv\langle a_m^{\dagger}a_n \rangle\nonumber
\ee 
The spectrum of $\bb C$ and $H_A$ have the following relation
\be
\nu_k=\frac{1}{e^{\eps_k}-1}\nonumber
\ee 
where $\{\nu_k\}$ is the spectrum of $\bb C$ and $\{\eps_k\}$ is spectrum of $H_A$. The von Neumann entropy and R\'enyi entropy are
\bea
S^{\rm vN}_A&=&-\tr{\rho_A\ln{\rho_A}}\\
&=&\sum_k\Big[\frac{\eps_k e^{-\eps_k}}{1-e^{-\eps_k}}-\ln(1-e^{-\eps_k}) \Big],\nonumber
\eea
and
\be
S^{{\rm R}_M}_A=\frac{\tr{\ln\rho_A^M}}{1-M}\nonumber
\ee 
Instead of using the spectrum $\{\eps\}$, one can use the spectrum $\{\nu_k\}$ to calculate von Neumann entropy and R\'enyi entropy as
\be
S^{\rm vN}_A=\sum_k (1+\nu_k)\ln(1+\nu_k)-\nu_k\ln\nu_k\nonumber
\ee 
and
\be
S^{{\rm R}_M}_A={\rm Tr}\ln\Big[ (1+\nu_k)^M- \nu_k^M\Big].\nonumber
\ee 
or equivalently in terms of the C matrix as
\be
S^{\rm vN}_A=-\tr{(\bb 1+\bb C)\ln(\bb 1+\bb C)-\bb C\ln \bb C}\nonumber
\ee 
and
\be
S^{{\rm R}_M}_A=-\frac{1}{1-M}\ln\det\Big[ (\bb 1+\bb C)^M- \bb C^M\Big].\nonumber
\ee 
\emph{Fermions} - The entanglement Hamiltonian is
\be
\rho_A=\frac{e^{-H_A}}{Z_A}\nonumber
\ee 
where $H_A=\sum_{mn} \eps_{mn}a_m^{\dagger}a_n$ and $Z_A=\prod_k (1+e^{-\eps_k})$. Again, we define the correlator
\be
C_{mn}\equiv\langle a_m^{\dagger}a_n \rangle\nonumber
\ee 
The spectrum of $\bb C$ and $H_A$ are related according to
\be
\nu_k=\frac{1}{e^{\eps_k}+1}\nonumber
\ee 
where $\{\nu_k\}$ is the spectrum of $\bb C$ and $\{\eps_k\}$ is spectrum of $H_A$. Therefore, von Neumann entropy and R\'enyi entropy are
\be
S^{\rm vN}_A=-\tr{\rho_A\ln{\rho_A}}=\sum_k\Big(\frac{\eps_k e^{-\eps_k}}{1+e^{-\eps_k}}+\ln(1+e^{-\eps_k}) \Big),\nonumber
\ee 
and
\be
S^{{\rm R}_M}_A=\frac{1}{1-M}\ln\tr{\rho_A^M}.\nonumber
\ee 
Instead of using the spectrum $\{\eps\}$, one can use the spectrum $\{\nu_k\}$ to calculate von Neumann entropy and R\'enyi entropy as
\be
S^{\rm vN}_A=-\sum_k (1-\nu_k)\ln(1-\nu_k)+\nu_k\ln\nu_k\nonumber
\ee 
and
\be
S^{{\rm R}_M}_A=tr \ln\Big[ (1+\nu_k)^M- \nu_k^M\Big]\nonumber
\ee 
which in terms of the $C$-matrix are given by
or
\be
S^{\rm vN}_A=-\tr{(1-\bb C)\ln(1-\bb C)+\bb C\ln \bb C}\nonumber
\ee 
and
\be
S^{{\rm R}_M}_A=\frac{1}{1-M}\ln\det\Big[ (\bb 1-\bb C)^M+\bb C^M\Big].\nonumber
\ee 
We can unify the entropies of Bosons and Fermions into the following
\be
S^{\rm vN}_A=\zeta\tr{(\bb 1+\tilde{\zeta}\bb C)\ln(\bb 1+\tilde{\zeta}\bb C)-\tilde{\zeta}\bb C\ln \bb C}\nonumber
\ee 
and
\be
S^{{\rm R}_M}_A=\frac{-\zeta}{1-M}\ln\det\Big[ (\bb 1+\tilde{\zeta}\bb C)^M-\tilde{\zeta}\bb C^M\Big].\nonumber
\ee 

\bw

\subsection{Replica symmetry and self-energy}\label{sec:replica}
In order to compute the R\'enyi entropy, one need to solve the large-N path-integral problem on an extended manifold shown in Fig. 1(c). The R\'enyi entropy is given by
\be
\tr{\rho_A^M}=\frac{1}{Z^M}\int{D[G,\Sigma]D\chi}e^{-N{\cal S}[\chi,G,\Sigma]}
\ee
here, $\chi$ represents the fermions and $G-\Sigma$ are used to decouple the interaction. Here, we show how this problem reduces to the action \pref{eq4} and \pref{eq6b} of the paper. In order to be concrete and without loss of generality, we consider the coupled SYK model \cite{Sohal2022}. This equilibrium path integral description of this model is reviewed in section \ref{sec:cSYK1} of the present Appendix.
The replica action is
\be
{\cal S}=\sum_{\mu=A,B}\sum_r\int d\tau \Bigg\{\frac{1}{2N}\sum_{i=1}^{N_{\mu}} \chi^{(r)}_{i,\mu}(\tau) \partial_{\tau} \chi^{(r)}_{i,\mu}(\tau)+\frac{J^\mu_{i,j,k,l}}{4!}\sum_{i,j,k,l=1}^{N_{\mu}}\chi^{(r)}_{i,\mu}\chi^{(r)}_{j,\mu}\chi^{(r)}_{k,\mu}\chi^{(r)}_{l,\mu}+\frac{V_{i,j,k,l}}{4!}\sum_{i,j=1}^{N_A}\sum_{k,l=1}^{N_B}\chi^{(r)}_{i,A}\chi^{(r)}_{j,A}\chi^{(r)}_{k,B}\chi^{(r)}_{l,B}\Bigg\}\\
\ee
which is diagonal in replica and needs to be supplemented with the boundary condition \pref{eq2}. The random variables $J^\mu_{ijkl}$ and $V^\mu_{ijkl}$ have zero mean and the variance 
\be
J^\mu_{ijkl} J^\nu_{i'j'k'l'}=\delta^{\mu\nu}\delta_{ii'}\delta_{jj'}\delta_{kk'}\delta_{ll'}\frac{J^2}{4N_{\mu}^3},\hspace{2mm}\langle V^2_{ijkl}\rangle =\delta_{ii'}\delta_{jj'}\delta_{kk'}\delta_{ll'}\frac{V^2}{(N_AN_B)^{\frac{3}{2}}}
\ee

After disorder averaging, the action develops off-diagonal-in-replica contributions and after $G-\Sigma$ decoupling becomes
\bea
{\cal S}&=&\sum_{r,r'} \sum_{\mu=A,B}\int d\tau d\tau'\Bigg\{\frac{1}{2N}\sum_{i} \chi^{(r)}_{i,\mu}(\tau) [\partial_{\tau}\delta(\tau,\tau')\delta^{rr'}+\Sigma^{(rr')}_{\mu}(\tau,\tau')] \chi^{(r')}_{i,\mu}(\tau')\\
&&\hspace{2cm}-\frac{1}{2} \Big[\Sigma^{(rr')}_{\mu}(\tau,\tau')G^{(r'r)}_{\mu}(\tau',\tau)+\frac{J^2}{4}[G^{(rr')}_{\mu}(\tau,\tau')]^4\Big]-\frac{V^2\sqrt{N_AN_B}}{2}G^{(rr')}_A(\tau,\tau')^2G^{(rr')}_B(\tau,\tau')^2 \Bigg\}\nonumber 
\eea
where $\mu$ and $\nu$ are $a$ or $b$ for self-energy $\Sigma_{\mu}$ and $A$ or $B$ for Green's function $G_{\mu}$. Note that the interacting part of the action contains inter-replica interaction and such four-fermion terms are decoupled by the $G^{(rr')}$ Green's function, leading to off-diagonal replica self-energy $\Sigma^{(rr')}$. Transforming from the replica sector  $n$, to replica momentum space $p$, we find
\bean
{\cal S}
&=& \frac{1}{2}\int d\tau d\tau' \Bigg\{ \frac{1}{N}\sum_{p,p'}\sum_{\mu=A,B}\sum_{i} \bar{\chi}^{(p)}_{i,\mu}(\tau) \left(\partial_{\tau}\delta(\tau,\tau')\delta^{pp'}+\Sigma^{(pp')}_{\mu}(\tau,\tau')\right) \chi^{(p')}_{i,\mu}(\tau')-\sum_{p,p'}\sum_{\mu}\Sigma^{(pp')}_{\mu}(\tau',\tau)G^{(pp')}_{\mu}(\tau,\tau')\\
&&-\sum_{p_1,p_2,p_3,p_1',p_2',p_3'}\Big[\frac{V^2\sqrt{N_AN_B}}{2M^2}G^{(p_1p'_1)}_A(\tau,\tau')G^{(p_2p'_2)}_A(\tau,\tau')G^{(p_3p'_3)}_B(\tau,\tau')G^{(-p_1-p_2-p_3,-p'_1-p'_2-p'_3)}_B(\tau,\tau') \\
&&+\frac{J^2}{4M^2}\sum_{\mu}G^{(p_1p_1')}_{\mu}(\tau,\tau')G^{(p_2p_2')}_{\mu}(\tau,\tau')G^{(p_3p_3')}_{\mu}(\tau,\tau')G^{(-p_1-p_2-p_3,-p_1'-p_2'-p_3')}_{\mu}(\tau,\tau')\Big]\Bigg\}\\
\eean
where we have used
\be
\chi^{(r)}(\tau)=\frac{1}{\sqrt
M}\sum_{p}\Omega^{-pr}\chi^{(p)}(\tau),\quad G^{(rr')}(\tau,\tau')=\frac{1}{M}\sum_{pp'}\Omega^{-(pr-p'r')}G^{(pp')}(\tau,\tau'),\quad \Sigma^{(rr')}(\tau,\tau')=\frac{1}{M}\sum_{pp'}\Omega^{-(pr-p'r')}\Sigma^{(pp')}(\tau,\tau'),\nonumber
\ee 
with inverse relations
\be
\chi^{(p)}(\tau)=\frac{1}{\sqrt M}\sum_{r}\Omega^{pr}\chi^{(r)}(\tau),\quad G^{(pp')}(\tau,\tau')=\frac{1}{M}\sum_{rr'}\Omega^{pr-p'r'}G^{(rr')}(\tau,\tau'),\quad \Sigma^{(pp')}(\tau,\tau')=\frac{1}{M}\sum_{rr'}\Omega^{(pr-p'r')}\Sigma^{(rr')}(\tau,\tau').\nonumber
\ee 
Now in the $p$ space, one set of saddle point solutions are found by varying the Green's function $G^{(pp')}_{\mu}$
\bean 
\Sigma^{(pp')}_{\mu}(\tau',\tau)&=&-\sum_{p_1,p_2,p_1',p_2'}\Big[\frac{J^2}{M^2}G^{(p_1p_1')}_{\mu}(\tau,\tau')G^{(p_2p_2')}_{\mu}(\tau,\tau')G^{(-p_1-p_2-p,-p_1'-p_2'-p')}_{\mu}(\tau,\tau')\\
&&+p^{\frac{\tilde{\mu}}{2}}\frac{V^2}{M^2}G^{(p_1p'_1)}_{\mu}(\tau,\tau')G^{(p_2p'_2)}_{-\mu}(\tau,\tau')G^{(-p_1-p_2-p,-p'_1-p'_2-p')}_{-\mu}(\tau,\tau')\Big]\nonumber
\eean
where $\tilde{\mu}$ is $\pm 1$ for $\mu=A$ and $B$ respectively. The $-\mu$ means the other part besides $\mu$. 
The variations w.r.t $\Sigma^{(pp')}_{\mu}$ gives the the Dyson equation
\be
G^{(p'p)}(\tau,\tau')=\frac{1}{N}\sum_i\braket{\chi_i^{(p)}(\tau')\chi^{(p')}_i(\tau)}={\cal G}_{p,p'}(\tau,\tau').
\ee
We note that while these equations do generally support replica off-diagonal solutions $G^{(pp')}$ and $\Sigma^{(pp')}$, However, replica symmetry is also preserved by these equations. This means if we assume $G^{(pp')}\propto\delta^{pp'}$ we find $\Sigma^{(pp')}\propto\delta^{pp'}$ which means the action decouples into different $p$ sectors, leading to $G^{(pp')}\propto\delta^{pp'}$. Considering that the at UV, $G\sim J\delta^{pp'}$ is replica symmetric, we conclude that the replica symmetry is preserved. Therefore, the Green's functions and self-energies can be represented by online diagonal $p$ indices, $G^{(p)}(\tau,\tau')={\cal G}_{p}(\tau,\tau')$. Likewise, the self consistency equations become the same for different $p$ sectors:
\ben
\Sigma^{(p)}_{\mu}(\tau',\tau)=-\sum_{p_1p_2}\Big[\frac{J^2}{M^2}G^{(p_1)}_{\mu}(\tau,\tau')G^{(p_2)}_{\mu}(\tau,\tau')G^{(-p-p_1-p_2)}_{\mu}(\tau,\tau')+p^{\frac{\tilde{\mu}}{2}}\frac{V^2}{M^2}G^{(p_1)}_{\mu}(\tau,\tau')G^{(p_2)}_{-\mu}(\tau,\tau')G^{(-p-p_1-p_2)}_{-\mu}(\tau,\tau')\Big]
\een 
The only thing different between various sectors is the boundary condition in imaginary-time direction. So, a full solution to the problem requires simultaneous solution to all $p$ sectors. Eventually, the $\tr{\rho^M_A}$ can be written as
\be
\tr{\rho^M_A}=\frac{1}{Z_0^M}\int{D\chi}e^{-N{\cal S}}, \qquad
{\cal S}={\cal S}_Q+{\cal S}_C
\ee
where using the saddle-point equations, the classical part is
\bean
{\cal S}_C&=&-\frac{1}{2}\int d\tau d\tau' \Big[\sum_p\sum_{\mu}\Sigma^{(p)}_{\mu}(\tau',\tau)G^{(p)}_{\mu}(\tau,\tau')+\frac{V^2\sqrt{N_AN_B}}{2M^2}\sum_{p_1,p_2,p_3}G^{(p_1)}_A(\tau,\tau')G^{(p_2)}_A(\tau,\tau')G^{(p_3)}_B(\tau,\tau')G^{(-p_1-p_2-p_3)}_B(\tau,\tau') \\
&&+\frac{J^2}{4M^2}\sum_{\mu}\sum_{p_1,p_2,p_3}G^{(p_1)}_{\mu}(\tau,\tau')G^{(p_2)}_{\mu}(\tau,\tau')G^{(p_3)}_{\mu}(\tau,\tau')G^{(-p_1-p_2-p_3)}_{\mu}(\tau,\tau')\Big]\\
\eean
and the quantum part of the action is
\be
{\cal S}_Q=\sum_{p}\sum_{\mu=A,B}\int d\tau d\tau'  \frac{1}{2N}\sum_{i} \bar{\chi}^{(p)}_{i,\mu}(\tau) \left(\partial_{\tau}\delta(\tau,\tau')+\Sigma^{(p)}_{\mu}(\tau,\tau')\right) \chi^{(p)}_{i,\mu}(\tau').
\ee

As we have argued in the paper, however, a full self-consistent solution to all $p$-sectors is not needed if we are only interested in the von Neumann entanglement entropy. In this case, we could assume that the self-energy $\Sigma^{(p)}(\tau,\tau')=\Sigma(\tau_1-\tau_2)$ has the same expression as the time-translational invariant $p=0$ sector.  Going to frequency space, the quantum action becomes
\bea
{\cal S}_Q&=&\frac{1}{2N}\sum_{p}\sum_i\sum_{n,m}
\left[\begin{array}{cc}
\bar{\chi}^{(p)}_{A,i}(i\bar{\omega}_n)    &\bar{\chi}^{(p)}_{B,i}(i\omega_m) 
\end{array}\right]
\left(
\begin{array}{cc}
   -i\bar{\omega}_n+ \Sigma^{nn}_{A}  & 0 \\
   0  &-i\omega_m + \Sigma^{mm}_{B}
\end{array}\right)
\left[\begin{array}{c}
  \chi^{(p)}_{A,i}(i\bar{\omega}_{n}) \\
  \chi^{(p)}_{B,i}(i\omega_{m})
\end{array}\right]
\eea
where  $n$ and indices for shifted Matsubara frequencies, different in each $p$ sector and $m$ are indices for normal Matsubara frequencies. The classical part becomes
\bea
{\cal S}_C&=&-\frac{3}{8} \sum_{p,n,m}\Big[\Sigma_{A}(i\bar{\omega}_n)G_{A}(i\bar{\omega}_n)+\Sigma_{B}(i\omega_m)G_{B}(i\omega_m)\Big].
\eea 
The elements of self-energies are worked out in section \ref{sec:action}. 

Then $G$ and $\Sigma$ will be just the equilibrium Green's functions and self-energies with $p=0$. 

\ew

\subsection{Construction of the action in the replica-momentum space}\label{sec:action}
In this section, we construct the elements of the matrix ${\cal G}_u^{-1}$ appearing in action \pref{eq4} of the paper. The diagonal elements are quite straight forward, so we focus on off-diagonal elements for both bosons and fermions. We use the following identities
\ben
\Sigma(\tau)=\frac{1}{\beta}\sum_n\Sigma(i\omega_n)e^{-i\omega_n\tau}, \quad \Sigma(z)=\int{\frac{dx}{\pi}}\frac{\Sigma''(x)}{x-z}.
\een
\emph{Fermions} - For the case of fermions we can write
\bea
\Sigma(\tau)&=&-\oint{\frac{dz}{2\pi i}}[f(z)-\theta_{\tau>0}]e^{-z\tau}\int{\frac{dx}{\pi}}\frac{\Sigma''(x)}{x-z}\nonumber\\
&=&\int{\frac{dx}{\pi}}[f(x)-\theta_{\tau>0}]\Sigma''(x)e^{-x\tau}
\eea
Note that $\theta_{\tau>0}$ is important for convergence, but also necessary to make sure that the Green's function is $\beta$ anti-periodic. The two-point version is simple but note that $\Sigma(\tau_1,\tau)=T\Sigma(\tau_1-\tau_2)$. Then,
\bea
\Sigma_{nm'}&=&T\intob{d\tau_1d\tau_2}e^{i(\bar\omega_n\tau_1-\omega_{m'}\tau_2)}\Sigma(\tau_1-\tau_2)
\eea
The result is
\bea
\Sigma_{nm'}&=&T\frac{1+u}{i\bar\omega_n-i\omega_{m'}}\int{\frac{dx}{\pi}}\frac{\Sigma''(x)}{i\omega_{m'}-x}\nonumber\\
&=&-T\frac{1+u}{i\bar\omega_n-i\omega_{m'}}\Sigma(i\omega_{m'}),
\eea
where $u=e^{i\bar\omega_n\beta}$. 
Note that choosing $\Sigma(\tau_1,\tau_2)=V\delta(\tau_1,\tau_2)$ or $\Sigma(i\omega_{m'})=V$ reproduces the known result:
\be
V\intob{d\tau e^{i(\bar\omega_n-\omega_{m'})\tau}}=-\frac{1}{\beta}V\frac{u+1}{i\bar\omega_n-i\omega_{m'}}.
\ee
Similarly, we can show that
\be
\Sigma_{mn'}=-T\frac{1+u^{-1}}{i\omega_m-i\bar\omega_{n'}}\Sigma(i\omega_m).
\ee
This is correct, because using $\bar\omega_n=\nu_n-iT\log u$ we find
\bean
\lim_{u\to -1}\Sigma_{nm'}&=&\lim_{\eps \to 0}\frac{-T\eps \delta_{nm'}\Sigma(i\omega_{m'})}{T\log(-1+\eps)-i\pi T}=\delta_{nm'}\Sigma(i\omega_{m'}),\\
\lim_{u\to -1}\Sigma_{mn'}&=&\lim_{\eps \to 0}\frac{-T[(-1+\eps)^{-1}+1]\delta_{mn'}\Sigma(i\omega_{m})}{i\pi T-T\log(-1+\eps)},\\
&&=\delta_{mn'}\Sigma(i\omega_{m}).
\eean
\emph{Bosons} - In this case we have
\bea
\Sigma(\tau)&=&\frac{1}{\beta}\sum_ne^{-i\nu_n\tau}\Sigma(i\nu_n)\nonumber\\
&=&\oint{\frac{dz}{2\pi i}}[n(z)+\theta_\tau]e^{-z\tau}\Sigma(z)\nonumber\\
&=&\int{\frac{dx}{\pi}}\Sigma''(x)[n(x)+\theta_\tau]e^{-x\tau}
\eea
This has the correct half-periodicity, as seen in
\bean
\Sigma(\beta-\abs{\tau})&=&\int{\frac{dx}{\pi}}\Sigma''(x)[n(x)+1]e^{-\beta x}e^{+x\abs{\tau}}\\
&=&\int{\frac{dx}{\pi}}\Sigma''(x)n(x)e^{+x\abs{\tau}}=\Sigma(-\abs{\tau})
\eean
Fourier transform is
\bean
\Sigma_{nm'}&=&T\int{\frac{dx}{\pi}\Sigma''(x)}\intob{d\tau_1d\tau_2}[n(x)+\theta_{\tau_1>\tau_2}]\\
&&\hspace{4.5cm} e^{i(
\bar\omega_n-x)\tau_1}e^{-(i\nu_{m'}-x)\tau_2}\\
&=&T\frac{u-1}{i\bar\omega_n-i\nu_{m'}}\int{\frac{dx}{\pi}\frac{\Sigma''(x)}{x-i\nu_{m'}}}=T\frac{u-1}{i\bar\omega_n-i\nu_{m'}}\Sigma(i\nu_{m'})
\eean
and
\bean
\Sigma_{mn'}&=&T\int{\frac{dx}{\pi}\Sigma''(x)}\intob{d\tau_1d\tau_2}[n(x)+\theta_{\tau_1>\tau_2}]\\
&&\hspace{4.5cm}e^{-i(
\bar\omega_{n'}-x)\tau_2}e^{(i\nu_{m}-x)\tau_1}\\
&=&T\frac{u^{-1}-1}{i\nu_m-i\bar\omega_{n'}}\int{\frac{dx}{\pi}\frac{\Sigma''(x)}{x-i\nu_m}}=T\frac{u^{-1}-1}{i\nu_m-i\bar\omega_{n'}}\Sigma(i\nu_m).
\eean\\

As a check, using $e^{i\beta\bar\omega_n}=u$ and $\bar\omega_n=\nu_n-iT\log u$ we find
\bean
\lim_{u\to 1}\Sigma_{mn'}=
\delta_{mn'}\Sigma(i\nu_{m}), \qquad \lim_{u\to 1}\Sigma_{nm'}=\delta_{nm'}\Sigma(i\nu_{m'})
\eean
So, in summary
\be
\Sigma_{nm'}=-\frac{1}{\beta}\frac{\tilde\zeta u-1}{i\bar\omega_n-i\omega_{m'}}, \qquad \Sigma_{mn'}=-\frac{1}{\beta}\frac{\tilde\zeta u^{-1}-1}{i\omega_{m}-i\bar\omega_{n'}}.\nonumber
\ee

\subsection{Useful matrix identities}\label{sec:identities}
In this section, we provide some useful matrix identities that are used in the paper. The first is the famous determinant identity
\bea
\det\left[
\begin{array}{c|c}
    A & B \\
    \hline
    C & D
\end{array}
\right]&=&\det(A)\det(D-BA^{-1}C)\\
&=&\det(D)\det(A-CD^{-1}B)\nonumber
\eea 
which leads to the equation employed in the paper:
\bea 
\det\left[
\begin{array}{c|c}
    I_{mm} & V_{mn} \\
    \hline
    V_{nm} & I_{nn}
\end{array} \label{deter}
\right]&=&\det(I_{mm}-V_{mn}V_{nm})\\ 
&=&\det(I_{nn}-V_{nm}V_{mn}).\nonumber 
\eea
We also use some matrix inversion identities. 
\bw

If $(A-BD^{-1}C)$ is invertible 
\be
\left[\begin{array}{cc}
    A & B \\
    C & D
\end{array}\right]^{-1}
=
\left[\begin{array}{cc}
    (A-BD^{-1}C)^{-1}&  -(A-BD^{-1}C)^{-1}BD^{-1} \\
    -D^{-1}C (A-BD^{-1}C)^{-1} & D^{-1}+D^{-1}C(A-BD^{-1}C)^{-1} BD^{-1}
\end{array}\right]
\ee 

If $(D-CA^{-1}B)$ is invertible
\be
\left[\begin{array}{cc}
    A & B \\
    C & D
\end{array}\right]^{-1}
=
\left[\begin{array}{cc}
    A^{-1}+A^{-1}B(D-CA^{-1}B)^{-1} CA^{-1}& -A^{-1}B(D-CA^{-1}B)^{-1} \\
    -(D-CA^{-1}B)^{-1}CA^{-1} & (D-CA^{-1}B)^{-1}
\end{array}\right]
\ee 

A consequence of these identities is that

\bea
V_{qk}G^B_{kk'}V_{k'q}&=&V_{qk}(G_b^{-1}-V_{kq}G_aV_{qk})^{-1}V_{kq}=
\mat{V_{qk}     &  0}
\mat{
  G_b^{-1}   & V_{kq} \\  V_{qk}   & G_a^{-1}  }
  \mat{     V_{kq} \\    0 }\nonumber\\
&=&V_{qk}[G_b+G_bV_{qk}(g_A^{-1}-V_{qk}G_bV_{kq})^{-1}V_{qk}G_b]V_{kq}\nonumber\\
&=&\Sigma_a+\Sigma_a G_A \Sigma_a=G_a^{-1}G_AG_a^{-1}-G_a^{-1}\label{matrix}
\eea

\subsection{A more detailed proof of Eq.\,\pref{eq7}}\label{sec:eq8}

Here, we provide a more detailed proof of central equation of the paper, Eq.\,\pref{eq7}. We start from the action \pref{eq4}

\bean
{\cal S}&=&\frac{1}{\beta}\sum_n \Big[\bar{\psi}_a(i\bar{\omega}_n)\Big(-i\bar{\omega}_n+\eps_a+\Sigma^{aa}(i\bar{\omega}_n)\Big)\psi_a(i\bar{\omega}_n)+\sum_m\bar{\psi}_b(i\omega_m)\Big(-i\omega_m+\eps_b+\Sigma^{bb}(i\omega_m)\Big)\psi_b(i\omega_m)\nonumber\\
&&\hspace{2cm}+\frac{1}{\beta^2}\sum_{n,m}\Big\{\Sigma^{ab}(i\omega_{m})\frac{e^{i\bar{\omega}_{n}\beta}-1}{i\bar{\omega}_{n}-i\omega_{m}}\bar{\psi}_a(i\bar{\omega}_{n}) \psi_b(i\omega_{m})+\Sigma^{ba}(i\omega_{m})\frac{e^{-i\bar{\omega}_{n}\beta}-1}{i\omega_{m}-i\bar{\omega}_{n}}\bar{\psi}_b(i\omega_{m}) \psi_a(i\bar{\omega}_{n})\Big\}\Big]\nonumber\\\nonumber
\eean

Shift $\psi_a(i\bar{\omega}_n) \to \psi_a(i\bar{\omega}_n)-\frac{1}{\beta}\sum_{m}\Sigma^{ba}(i\omega_{m}) \frac{1}{i\bar{\omega}_{n}-\eps_a-\Sigma_{aa}(i\bar{\omega}_{n})}\frac{1-u^{-1}}{i\bar{\omega}_{n}-i\omega_{m}} \psi_b(i\omega_{m})$, the action becomes

\bea 
{\cal S}&=&{\cal S}_a+\frac{1}{\beta^2}\sum_{m,m',n}\bar{\psi}_b(i\omega_{m})\Big[-[G^{-1}(i\omega_{m})]^{bb'}\delta_{m,m'}\nonumber\\
&&\hspace{2cm}+\frac{1}{\beta}\Sigma^{ba}(i\omega_{m}) \frac{1}{i\bar{\omega}_{n}-\eps_a-\Sigma^{aa'}(i\bar{\omega}_{n})}\frac{(1-u)(1-u^{-1})}{(i\bar{\omega}_{n}-i\omega_{m})(i\bar{\omega}_{n}-i\omega_{m'})}\Sigma^{a'b'}(i\omega_{m'}) \Big]\psi_b(i\omega_{m'})\nonumber\\\nonumber
\eea 
where $[G^{-1}(i\omega_{m})]^{bb'}=-i\omega_{m}+\eps_b+\Sigma^{bb'}(i\omega_{m})$.  Using spectral representation
\be
{\cal S}={\cal S}_a+\frac{1}{\beta^2}\sum_{m,m'}\bar{\psi}_b(i\omega_{m})\Big[-[G^{-1}(i\omega_{m})]^{bb'}\delta_{m,m'}+\frac{1}{\beta}\sum_{n}\Sigma^{ba}(i\omega_{m}) \int \frac{dx}{2\pi}\frac{A^{aa'}(x)}{i\bar{\omega}_{n}-x}\frac{(1-u)(1-u^{-1})}{(i\bar{\omega}_{n}-i\omega_{m})(i\bar{\omega}_{n}-i\omega_{m'})}\Sigma^{a'b'}(i\omega_{m'}) \Big]\psi_b(i\omega_{m'})\nonumber\\
\ee

\ew

After integrating out the shifted Matsubara frequency, the action becomes

\bea
{\cal S}&=&{\cal S}_a+\frac{1}{\beta^2}\sum_{m,m'}\bar{\psi}_b(i\omega_{m})
\Big[-[G^{-1}(i\omega_{m})]^{BB'}\delta_{m,m'}\\
&&+\int{\frac{dx}{2\pi}}\frac{K_u(x)\Sigma^{ba}(i\omega_m)A^{aa'}(x)\Sigma^{a'b}(i\omega_{m'})}{(i\omega_{m}-x)(i\omega_{m'}-x)}\Big]
\psi_b(i\omega_{m'})
\nonumber
\eea

where $-[G^{-1}(i\omega_{m})]^{BB'}$ is
\be
-[G^{-1}(i\omega_{m})]^{bb'}+\int\frac{dx}{2\pi}\frac{\Sigma^{ba}(i\omega_{m})A^{aa'}(x)\Sigma^{a'b'}(i\omega_{m})}{i\omega_{n}-x}\nonumber
\ee

Then the whole second part in action $S$ gives $({\cal G}_u^{BB'})^{-1}$ which is

\be
[G^{-1}_m]^{BB'}\delta_{mm'}-\int{\frac{dx}{2\pi}} \frac{K_u(x)\Sigma^{ba}_m A^{aa'}(x)\Sigma^{a'b'}_{m'}}{\beta(i\omega_m-x)(i\omega_{m'}-x)}\nonumber
\ee

The determinant $\det{}^{-\zeta}[(-{\cal G}_u^{BB'})^{-1}]$ now is
\be
\det{}^{-\zeta}\Big[[G^{-1}_m]^{BB'}\delta_{mm'}-\int{\frac{dx}{2\pi}} \frac{K_u(x)\Sigma^{ba}_m A^{aa'}(x)\Sigma^{a'b'}_{m'}}{\beta(i\omega_m-x)(i\omega_{m'}-x)} \Big]\nonumber
\ee

Now we are facing a determinant
\be 
\det{}^{-\zeta}\Big[D_{mm'}\delta_{mm'}+\sum_x w_{m}(x)v^T_{m'}(x)\Big]\nonumber
\ee 

This is nothing but
\begin{figure}[h!]
\includegraphics[width=0.75\linewidth]{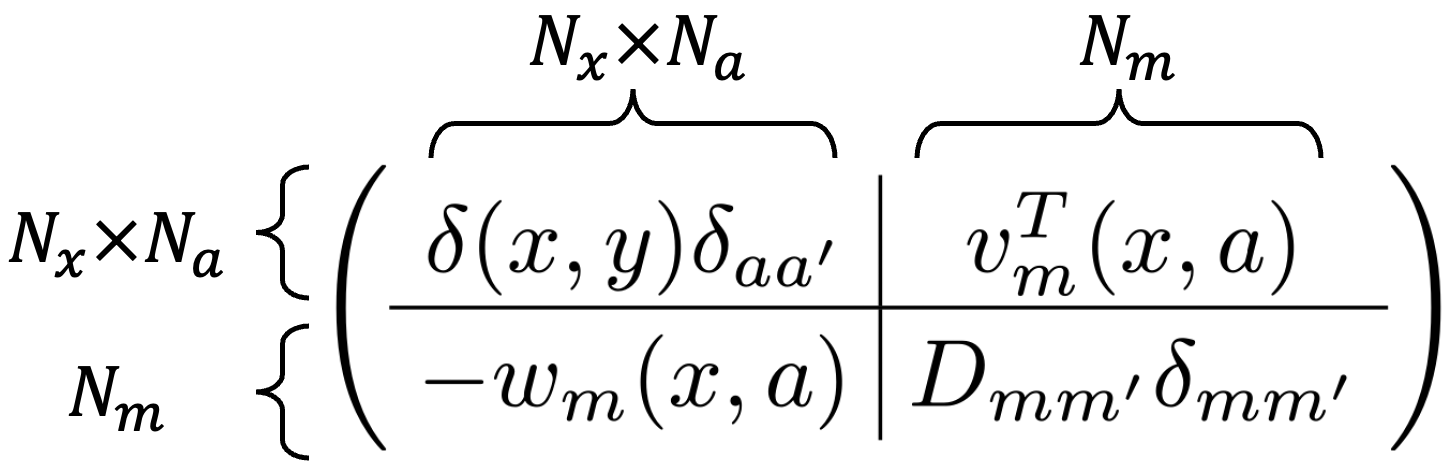}
\end{figure}

\noindent where $N_x$, $N_a$ and $N_m$ refer to the number of real frequencies, number of A modes and number of Matsubara frequencies, respectively. Then the determinant is
\be 
\det{}^{-\zeta}\Big[\delta(x,y)\delta_{aa'}+\sum_m v^T_{m}(x,a)D_{mm}w_{m}(y,a')\Big]\nonumber
\ee 
Using Eq.\,\pref{deter}, and after the shuffling the determinant becomes
\bean
&&\det{}^{-\zeta}\Big[\delta(x,y)\\
&&\hspace{1cm}-\frac{1}{\beta}\sum_m K_u(x)A^a(x)\frac{\Delta x}{2\pi}\frac{\Sigma_m^{ab}G_m^{BB'}\Sigma_m^{ba}}{(i\omega_{m}-x)(i\omega_{m}-y)}\Big]\nonumber
\eean
This motivates defining [$z$ is the complex frequency]
\be
\bb R(z)\equiv\Sigma^{ab}(z)G^{BB'}(z)\Sigma^{ba}(z)
\ee
in terms of which
\bea
\bb r(x)&\equiv&\frac{1}{\beta}\sum_n\frac{\bb R(i\omega_n)}{i\omega_n-x}\\
&=&\int\frac{d\omega}{2\pi}\bb A_R(\omega)\frac{n_{\tilde\zeta}(x)-n_{\tilde\zeta}(\omega)}{x-\omega}.\label{eq36c}
\eea
Here, we have used the spectral representation of the $R(z)$ function, defined as $\bb A_R(\omega)=i[\bb R(\omega+i\eta)-\bb R(\omega-i\eta)]$. We also define
\be
\bb J(x,y)\equiv\frac{\bb r(x)-\bb r(y)}{x-y}\label{eq37c}
\ee
Note that the $x\to\omega$ limit in Eq.\,\pref{eq36c} and $x\to y$ limit in Eq.\,\pref{eq37c} needs to be treated using L'H\^opital's rule. In terms of the $\bb J$ matrix we find
\be
\det{}^{-\zeta}[\bb 1+\tilde\zeta K_u(\omega)\bb A^a(\omega)\bb J^A(\omega,\omega')],\nonumber
\ee  
Finally, we get the $u$-sector partition function \pref{eq7}
\be
Z[u]= {Z_a[u]}{Z_B}\det{}^{-\zeta}[\bb 1+\tilde\zeta K_u(\omega)\bb A^a(\omega)\bb J^A(\omega,\omega')].
\ee

\subsection{Non-interacting limit of our formalism}\label{sec:non-int-limit}

In this part, we will show that our approach can be connected to results in the non-interacting limit. In the non-interacting case, the self-energy in the action \pref{eq4} is just a frequency independent constant $V$. Thus, 
\bean
A_R^{aa'}(\omega)&=&\sum_{b,b'}\text{Im}\Big[V^{ab}G^{BB'}(\omega+i\eta)V^{b'a'}\Big]\\
&=&\sum_{b,b'}V^{ab}A^{BB'}V^{b'a'}
\eean
In the non-interacting limit, the spectral function is 
\be
A^{aa'}(\omega)=2\pi\sum_{\eps}\phi_{\eps}(a)\phi^{*}_{\eps}(a')\delta(\omega-\eps)\nonumber
\ee
, so that the last determinant in \pref{eq7} reduces to
\be
\det{}^{-\zeta}\Big[\phi^{\dagger}_{\eps}(a') \Big(\delta(x,y)-\tilde{\zeta}K_u(x)\delta(x-\eps)J^{a'a}(x,y)\Big)\phi_{\eps}(a)\Big]\nonumber
\ee 
The $\phi_{\eps}(a)$ and $\phi^{\dagger}_{\eps}(a)$ in the determinant plays the role of unitary transformation from $a$ modes to $\eps$ modes which together with the $\delta(x-\eps)$ reduces the dimension of the determinant:
\be 
\det{}^{-\zeta}\Big[\delta(\eps,\eps')-\tilde{\zeta}K_u(\eps)J^{\eps\eps'}(\eps,\eps')\Big]\label{det1}
\ee  
where $J(\eps,\eps')$ can be written as
\be
J(\eps,\eps')=\frac{r(\eps)-r(\eps')}{\eps-\eps'},
\ee
in terms of $r(\eps)$
\be
r(\eps)=-\int\frac{d\omega}{2\pi}{A_R(\omega)}\frac{n_{\tilde{\zeta}}(\eps)-n_{\tilde{\zeta}}(\omega)}{\eps-\omega}.
\ee
The diagonal terms $J(\eps,\eps')$ need to be treated in a limiting procedure. In this context, $J(\eps,\eps')$ becomes
\be
J(\eps,\eps')=-\int\frac{d\omega}{2\pi}\frac{A_R(\omega)}{\eps-\eps'}\Big[\frac{n_{\tilde{\zeta}}(\eps)-n_{\tilde{\zeta}}(\omega)}{\eps-\omega}-(\eps\to\eps')\Big]\nonumber
\ee
With the help of $VG_BV=G_a^{-1}(G_A-G_a)G_a^{-1}$ which is shown in \pref{matrix}, we can further show
\bea
V^{ab}A^{BB'}V^{b'a'}
&=&iV^{ab}(G^{BB'}_{R}-G^{BB'}_{Ad})V^{b'a'}\nonumber\\
&=&(\omega-\eps)A^{AA'}(\omega)(\omega-\eps')\nonumber
\eea
So that $J(\eps,\eps')$ becomes
\bean
J(\eps,\eps')&=&\int\frac{d\omega}{2\pi}A^{AA'}(\omega)n_{\tilde{\zeta}}(\omega)+\delta(\eps,\eps')\frac{n_{\tilde{\zeta}}(\eps)\eps'-n_{\tilde{\zeta}}(\eps')\eps}{\eps-\eps'}\\
&&\hspace{1cm}-\int\frac{d\omega}{2\pi}\omega A^{AA'}(\omega)\frac{n_{\tilde{\zeta}}(\eps)-n_{\tilde{\zeta}}(\eps')}{\eps-\eps'}\nonumber\\
&=&-\tilde{\zeta}\frac{1}{\beta}\sum_n e^{-i\omega_n 0^{-}}G_A(i\omega_n)\\
&&\hspace{1cm}+n'_{\tilde{\zeta}}(\eps)\eps-n_{\tilde{\zeta}}(\eps)-n'_{\tilde{\zeta}}(\eps)\eps\nonumber\\
&=& -\tilde{\zeta}G_A(\tau=0^{-})-n_{\tilde{\zeta}}(\eps)\nonumber\\
&=& \langle a^{\dagger} a \rangle-n_{\tilde{\zeta}}(\eps)\nonumber
\eean
where we use $\int\frac{d\omega}{2\pi}\omega A^{AA'}(\omega)=\eps\delta(\eps,\eps')$. And we can further define $\bb C=\langle a^{\dagger} a \rangle$ for both Bosons and Fermions, the $\tr{\rho_A^M}$ becomes
\bea
\tr{\rho_A^M}
&=&\prod_u \frac{Z_a(u)}{Z_a} \det{}^{-\zeta}\Big[\delta(\eps,\eps')-\tilde{\zeta}K_u(\eps)(\bb C-n_{\tilde{\zeta}}(\eps))\Big]\label{det1}\nonumber\\
&=&\det{}^{-\zeta}\Big[\bb C^M+\bb D^M\Big]\nonumber
\eea
where  $\bb D=\bb 1+\tilde{\zeta}\bb C=\langle a a^{\dagger}\rangle$. The term $\prod_u [{Z_a(u)}/{Z_a}]$ can be just absorbed into the determinant to get the final expression because the dimension of the determinant in non-interacting case is finite, but one don't have this luxury again in the interacting case.  Finally, the von Neumann entropy and R\'enyi entropy are
\be
S^{\rm vN}_A=\tilde{\zeta}\tr{\bb D\ln \bb D+\tilde{\zeta}\bb C\ln \bb C}\nonumber
\ee
and
\be
S^{{\rm R}M}_A=\frac{-\zeta}{1-M}\ln\det\Big[\bb D^M+\tilde{\zeta}\bb C^M\Big]\nonumber
\ee 
which are exactly Casini's results using the reduced density matrix method.

\subsection{The perturbative limit of our formalism}\label{sec:perturbative}
The EE $S_A=S_a+\Delta S_A$ has a thermal part $S_a$ and a quantum correction. The latter can be expressed as
\be
\Delta S_A=\int{dc}\Delta\rho(c)g^{{\rm R}_M}(c)
\ee
 is expressed in terms of 
 \be
g^{{\rm R}_M}(c)\equiv\frac{-\zeta}{1-M}\log[(1+\tilde\zeta c)^M-\tilde\zeta c^M],
 \ee
 and the entanglement density of states
 \be
\Delta\rho(c)=-\frac{1}{\pi}\partial_c\im{{\rm Tr}\log\{(c^+\bb 1-\bb C_0)^{-1}(c^+\bb 1-\bb C)\}}.\nonumber
\ee
We can write the $\bb C$ matrix as $\bb C=\bb C_0+\bb \Pi$ where 
\be
\bb C_0=\tilde\zeta n_{\tilde\zeta}(\omega)\bb 1,\qquad
\bb\Pi=\sqrt{\bb A_a(\omega)}\bb J_A(\omega,\omega')\sqrt{\bb A_a(\omega')}.
\ee
In the perturbative the entanglement density of states is given by
\be
\Delta\rho(c)=-\frac{1}{\pi}\partial_c\im{{\rm Tr}\log\{\bb 1-(c^+\bb 1-\bb C_0)^{-1}\bb\Pi\}}
\ee
This expression can be expanded perturbatively. The leading order term is
\be
\Delta\rho(c)\approx\frac{1}{\pi}\partial_c\int{\frac{d\omega}{2\pi}}\im{\frac{{\rm Tr}\bb\Pi(\omega,\omega)}{c+i\eta-\tilde\zeta n(\omega)}}\\
\ee
${\rm Tr}[\bb\Pi(\omega,\omega)]$ is real. Therefore.
\be
\Delta\rho(c)=-\partial_c\int{\frac{d\omega}{2\pi}}\delta[c-\tilde\zeta n_{\tilde\zeta}(\omega)]{\rm Tr}\bb\Pi(\omega,\omega)
\ee
Note that in the limit $x\to y$ from Eq.\,\pref{eq37c} we find $\bb J(\omega,\omega)=\partial_\omega \bb r(\omega)$.
Instead the $\omega$-integral, next we do the $c$-integral and find
\be
\Delta S^{{\rm R}_M}_A=-\int{\frac{d\omega}{2\pi}}\tilde g^{{\rm R}_M}(\omega)\tr{\bb A^a(\omega)\partial_\omega\bb r(\omega)}\label{eq41b}
\ee
where $\tilde g^{{\rm R}_M}(\omega)$ is defined as
\be
\tilde g^{{\rm R}_M}(c)\equiv\partial_c g^{{\rm R}_M}(c)
\ee
and using $c=\tilde\zeta n_{\tilde\zeta}(\omega)=1/(e^{\beta\omega}-\tilde\zeta)$ we can express it as
\be
\tilde g^{{\rm R}_M}(\omega)=\zeta\frac{M}{M-1}\frac{n_{\tilde\zeta}(M\omega)}{n_{\tilde\zeta}[(M-1)\omega]n_{\tilde\zeta}(\omega)}.
\ee
Note that this expression has a well-behaved limit as $M\to 1^+$. The leading perturbative quantum correction to the R\'enyi entropy, Eq.\,\pref{eq41b}, can alternatively be extracted from Eq.\,\pref{eq7}. Perturbation theory in $J(\omega,\omega')$ gives
\bea
S_A^{\rm R_M}&=&\frac{1}{1-M}\log\frac{Z_M}{Z_0^M}\\
&=&\frac{\abs{\zeta}}{M-1}\int{\frac{d\omega}{2\pi}}\tr{\bb A^a(\omega)\partial_\omega\bb r(\omega)}\sum_pK_p(\omega)\nonumber
\eea
 $\sum_pK_{u_p}(\omega)$ can be computed using contour integration with $\partial_u\log(u^M-\tilde\zeta)$, which gives
 \be
\sum_pK_{u_p}(\omega)=\tilde\zeta M\frac{n_{\tilde\zeta}(M\omega)}{n_{\tilde\zeta}(\omega)n_{\tilde\zeta}[(M-1)\omega]}
 \ee
 which gives the same result.
 
\bw

\subsection{Review of equilibrium path integral of the coupled-SYK model}\label{sec:cSYK1}
For the sake of completeness, we list the saddle point equations and thermal dynamical properties of the coupled-SYK model. While it has been shown that such models can undergo a spontaneous symmetry breaking \cite{Kim2019}, we restrict our analysis to symmetry-preserving phases. We provide the self-consistency equations in the real-frequency so that they can be readily used with our real-frequency formalism.

The Hamiltonian of this coupled SYK model is
\bean
H&=&\frac{1}{4!}\sum_{\mu=A,B}\sum_{i,j,k,l}^{N_{\mu}}J_{ijkl}^{\mu} \chi_i^{\mu}\chi_j^{\mu}\chi_k^{\mu}\chi_l^{\mu}+\sum_{i,j=1}^{N_A}\sum_{k,l=1}^{N_B}V_{ijkl}\chi_i^A\chi_j^A\chi_k^B\chi_l^B \nonumber
\eean
and  $J_{ijkl}^{\mu}$,  $V_{ijkl}$ are Gaussian random variables with zero mean and variances
\ben
\langle (J_{ijkl}^{\mu})^2 \rangle=\frac{3!J^2}{N_{\mu}^3},\qquad\langle V_{ijkl}^2 \rangle=\frac{V^2}{(N_AN_B)^{\frac{3}{2}}}
\een
In this case, the action is
\bean 
{\cal S}&=& -\sum_{\mu} N_{\mu}\ln Pf(\partial_{\tau}+\Sigma_{\mu})-\int d\tau d\tau'\sum_{\mu}\frac{N_{\mu}}{2}(\Sigma_{\mu}(\tau,\tau')G_{\mu}(\tau',\tau)+\frac{J^2}{4}G_{\mu}(\tau,\tau')^4)-\frac{V^2\sqrt{N_AN_B}}{2}G_A(\tau,\tau')^2G_B(\tau,\tau')^2 
\eean
\emph{Self-consistent equations} - In the large-N limit, we can get the saddle point solutions 
\be
\Sigma_a(\tau,\tau')
=J^2G_A^3(\tau,\tau')+2V^2\sqrt{p}G^2_B(\tau,\tau')G_A(\tau,\tau'),\qquad
\Sigma_b(\tau,\tau')
=J^2G_B^3(\tau,\tau')+2V^2\sqrt{\frac{1}{p}}G^2_A(\tau,\tau')G_B(\tau,\tau')\nonumber
\ee
where $p={N_B}/{N_A}$. 

 These relations can be brought to real frequency by introducing $B_{\mu}(\tau,\tau')=G^2_{\mu}(\tau,\tau')$. Generally, we have the symmetries $G_{\mu}(\tau_1,\tau_2)=-G_{\mu}(\tau_2,\tau_1)$ which imply
$G_{\mu}(i\omega_n)=-G_{\mu}(-i\omega_n)$ in Matsubara frequency. In the Matsubara frequency domain, they become
\be
\Sigma_{a}(i\omega_n)=\frac{J^2}{\beta^2}\sum_{n_1,n_2}G_{A}(i\omega_{n_1})G_{A}(i\omega_{n_2})G_{A}(i\omega_{n}-i\omega_{n_1}-i\omega_{n_2})+\frac{2V^2\sqrt{p}}{\beta^2}\sum_{n_1,n_2}G_{B}(i\omega_{n_1})G_{B}(i\omega_{n_2})G_{A}(i\omega_{n}-i\omega_{n_1}-i\omega_{n_2}),\nonumber
\ee
and
\be
\Sigma_{b}(i\omega_n)=\frac{J^2}{\beta^2}\sum_{n_1,n_2}G_{B}(i\omega_{n_1})G_{B}(i\omega_{n_2})G_{B}(i\omega_{n}-i\omega_{n_1}-i\omega_{n_2})+\frac{2V^2\sqrt{\frac{1}{p}}}{\beta^2}\sum_{n_1,n_2}G_{A}(i\omega_{n_1})G_{A}(i\omega_{n_2})G_{B}(i\omega_{n}-i\omega_{n_1}-i\omega_{n_2}),\nonumber
\ee
We define spectral bosonic and fremions spectral functions $ A_{\mu}^B(\omega)$ and $A_{\mu}^G(\omega)$ as
\be
A_{\mu}^G(\omega)\equiv i[G_{\mu}(\omega+i\eta)-G_{\mu}(\omega-i\eta)], \andd  A_{\mu}^B(\omega)\equiv i[ B_{\mu}(\omega+i\eta)-B_{\mu}(\omega-i\eta)].
\ee
When analytically continued onto the real frequency axis, the self energies become
\bean
\Sigma_{a}(\omega+i\eta)&=&J^2\int \frac{d\omega'}{2\pi}[A^G_{A}(\omega')f(\omega')B_{A}(\omega+i\eta-\omega')+G_{A}(\omega+i\eta-\omega')n_B(-\omega')A^B_{A}(\omega)]\\
&&+2V^2\sqrt{p}\int \frac{d\omega'}{2\pi}[A^G_{A}(\omega')f(\omega')B_{B}(\omega+i\eta-\omega')+G_{A}(\omega+i\eta-\omega')n_B(-\omega')A^B_{B}(\omega)],
\eean
\bean
\Sigma_{b}(\omega+i\eta)&=&J^2\int \frac{d\omega'}{2\pi}[A^G_{B}(\omega')f(\omega')B_{B}(\omega+i\eta-\omega')+G_{B}(\omega+i\eta-\omega')n_B(-\omega')A^B_{B}(\omega)]\\
&&+2V^2\sqrt{\frac{1}{p}}\int \frac{d\omega'}{2\pi}[A^G_{B}(\omega')f(\omega')B_{A}(\omega+i\eta-\omega')+G_{B}(\omega+i\eta-\omega')n_B(-\omega')A^B_{A}(\omega)],
\eean
where
\ben
B_{\mu}(\omega+i\eta)=\int \frac{d\omega'}{2\pi}A_{\mu}(\omega')G_{\mu}(\omega+i\eta-\omega')[f(\omega')-f(-\omega')]
\een

\emph{Free energy and thermal entropy} - The free energy is equal to
\be
\begin{aligned}
\frac{\beta F}{N}
=&-\frac{1}{1+p}\ln Pf (\partial_{\tau}\delta(\tau,\tau')+\Sigma_{a}(\tau,\tau'))-\frac{1}{2(1+p)}\int d\tau d\tau'\frac{3}{4}\Sigma_a(\tau,\tau')G_A(\tau',\tau)\\
&-\frac{1}{1+1/p}\ln Pf (\partial_{\tau}\delta(\tau,\tau')+\Sigma_{b}(\tau,\tau'))-\frac{1}{2(1+1/p)}\int d\tau d\tau'\frac{3}{4}\Sigma_b(\tau,\tau')G_B(\tau',\tau)
\end{aligned}
\ee 
In Matsubara frequency domain and analytically continued onto the real frequency axis, the free energy density becomes
\ben
\frac{F}{N}=\int \frac{d\omega}{\pi} \Bigg\{\frac{1}{2}f(\omega)\text{Im}\sum_{\mu}\frac{1}{1+\zeta^p_{\mu}}\Big[\log (-\omega-i\eta+\Sigma_{\mu})+\frac{3}{4}\Sigma_{\mu}(\omega+i\eta)G_{\mu}(\omega+i\eta)\Big]\Bigg\},
\een
where $\zeta^p_{\mu}=p$ for $\mu=A$, $1/p$ for $\mu=B$. The thermal entropy is
\bean
\frac{S}{N}&=&-\frac{1}{2}\int \frac{d\omega}{\pi} \Bigg\{\partial_T f(\omega)\text{Im}\sum_{\mu}\frac{1}{1+\zeta^p_{\mu}}\Big[\log (-\omega-i\eta+\Sigma_{\mu})+\Sigma_{\mu}(\omega+i\eta)G_{\mu}(\omega+i\eta)\Big]\\
&&-\frac{J^2}{4}\sum_{\mu}B_{\mu}(-\omega-i\eta)\Big[\partial_T n_B(\omega)B_{\mu}(\omega+i\eta)+\Big(n_B(\omega)-n_B(-\omega)\Big)\partial_T B_{\mu}(\omega+i\eta)\Big]\\
&&-\frac{V^2\sqrt{p}}{1+p}B_{A}(-\omega-i\eta)\Big[\partial_T n_B(\omega)B_{B}(\omega+i\eta)+\Big(n_B(\omega)-n_B(-\omega)\Big)\partial_T B_{B}(\omega+i\eta)\Big]\Bigg\}
\eean

\ew

\end{document}